\documentclass[preprint,aps,prb,epsf]{revtex4}   % power law ref if prb
\include{amsmath}
\include{epsfx}
\usepackage{graphicx}
\begin{document}
\title{{\bf Influence of Hydrodynamic Interactions on Mechanical Unfolding of Proteins}}
\author{{\bf P. Szymczak $^1$ and Marek Cieplak$^2$}}

\address{
$^1$Institute of Theoretical Physics, Warsaw University,
ul. Ho\.za 69, 00-681 Warsaw, Poland\\
$^2$Institute of Physics, Polish Academy of Sciences,
Al. Lotnik\'ow 32/46, 02-668 Warsaw, Poland}

\begin{abstract}
We incorporate hydrodynamic interactions in a structure-based model of
ubiquitin and demonstrate that the
hydrodynamic coupling may reduce the peak force when stretching the
protein at constant speed, especially at larger speeds.
Hydrodynamic interactions are also shown to facilitate
unfolding at constant force and inhibit stretching by fluid flows.
\end{abstract}
\maketitle

\section{Introduction}

It has been widely recognized that the water environment
affects the energy landscape and functionality of biomolecules in a profound way.
There is, however, another solvent-related effect that is considered less frequently:
the hydrodynamic interactions (HI) between individual
segments of a biomolecule that moves.
These interactions may affect the dynamics of conformational changes because
any motion of one segment generates a local fluid flow which influences another
segment.
%This coupling between the dynamic
%degrees of freedom, mediated by the solvent, is commonly known as ``hydrodynamic
%interactions'' (HI).

The presence of HI is known to affects dynamic properties of soft mater.
For instance, HI modify the values of diffusion coefficients in colloidal suspensions
\cite{Dhont}, affect the characteristics of the coil-stretch transition in polymers
\cite{Larson}, change the kinetic pathways of phase separation in binary mixtures
\cite{Tanaka}, and
alter the kinetics of macromolecule adsorption on surfaces \cite{Wojtaszczyk}.
Much less is known about the role of HI in protein folding and unfolding processes.
Dickinson \cite{Dickinson} and Tanaka \cite{Tanaka2} speculated that HI
might affect the kinetics of protein folding, but the actual numerical
assessment of the role of HI has come with the paper by
Baumketner and Hiwatari \cite{Baumketner}. They have considered
coarse grained models and found that
HI delay folding of a $\beta$ hairpin but do not affect folding
of the $\alpha$-helix.

In this paper, we consider mechanical stretching of proteins and
study the relevance of HI to the process. The stretching can be accomplished
in several ways and we discuss three modes: at constant speed, at constant force,
and through fluid flow.
We chose ubiquitin as a model system, since there is a large body of experimental
\cite{Vazquez,Chyan,FernandezLi,Schlierf}
 and theoretical \cite{Makarov,Makarov2,Irback,cieplakmarszalek,cs1,West} data on its
unfolding. A coarse-grained, Go-type model \cite{Goabe} of a protein is used, constructed
based on the knowledge of its native state. The Go models have been shown to give
surprisingly good agreement with both the experimental results
\cite{ciepho,cieplakmarszalek,sulkowska}
and all-atom molecular dynamic simulations \cite{West} when it comes to stretching.

We outline the model in section 2, then introduce two different ways of tracking the
evolution of the system: through Langevin Dynamics
and Brownian Dynamics in sections 3 and 4 respectively.
In the following sections we discuss results pertaining to the three
modes of stretching and demonstrate that HI can take many roles: they inhibit
unfolding by fluid flow, but make the constant force stretching faster.
At constant speed, they reduce the peak force if the speed is sufficiently high.
This HI-related reduction in force may be downplayed in the all-atom simulations
of titin by Lu and Schulten \cite{Schulten} which would provide
part of an explanation
for the excessively large forces obtained in these studies.

\section{The coarse-grained protein model}

In our simulations, we use the coarse-grained, Go-type model of a protein.
In the model, each residue is represented by a single bead centered on the
position of the C$^{\alpha}$ atom. The successive beads along the backbone are
tethered by harmonic potentials with a minimum at 3.8 {\AA}. The other interactions
between the residues are split into two classes: native and non-native. This determination
is
made by
checking for native overlaps of all atoms in aminoacids when
represented by enlarged van der Waals spheres as proposed in
reference \cite{Tsai}. The amino acids, $i$ and $j$, that do overlap in
this sense are endowed with the effective Lennard-Jones potential
$V_{ij} = 4\epsilon \left[ \left( \frac{\sigma_{ij}}{r_{ij}}
\right)^{12}-\left(\frac{\sigma_{ij}}{r_{ij}}\right)^6\right]$.
The length parameters $\sigma _{ij}$ are chosen so that the potential
minima correspond, pair-by-pair, to the experimentally established
native distances between the respective aminoacids.
In order to prevent emergence of
entanglements, the non-native contacts are endowed with
a hard core repulsion described by the $r_{ij}^{-12}$ part of the Lennard-Jones potential
combined with a constant shift term
that makes the potential vanish smoothly at 4 {\AA}.
The specificity of a protein is contained in
the length parameters $\sigma _{ij}$.
The energy parameter, $\epsilon$,
is taken to be uniform. We take $\epsilon/k_B= 900 K$, which correlates
well with the data on titin and ubiquitin unfolding  \cite{Pastore,cieplakmarszalek}.
Thus the reduced temperature, $\tilde{T}=k_BT/\epsilon \;$ of 0.3
should be close to room temperature. All of the simulations
reported here were performed at this temperature.
Various simulation methods to study the dynamics of the system are
 outlined in the following sections.

\section{Langevin dynamics (LD)}

In this case, the dynamics of a protein is assumed to be governed
 by the Langevin equation
\begin{equation}
m \ddot{{\bf r_i}} = -\gamma (\dot{{\bf r}}_i-{\bf u}({\bf r}_i)) +
{\bf F}^c_i + {\bf \Gamma}_i \;\;.
\label{lang}
\end{equation}
Here, ${\bf r_i}$ is the position of the $i$'th aminoacid, ${\bf F}^c_i$ is the
net force on it due to contact potentials, $\gamma$ is the friction
coefficient, and ${\bf u}({\bf r}_i)$ denotes the solvent
flow field. Finally, ${\bf \Gamma}$ is a white noise term with the dispersion obeying
\begin{displaymath}
\langle {\bf \Gamma}_i(t) {\bf \Gamma}_j(t^\prime) \rangle = 2 k_BT \gamma \delta(t-t^
\prime) {\bf I} \delta_{ij}
\end{displaymath}
where {\bf I} is the identity matrix.
The white noise term mimicks the effect of the random collisions of the
aminoacids with the surrounding solvent at the same time serving as a thermostat of
the system. However, this scheme completely neglect the effects of HI which
may exist in a real system, when the motion of one particle induces the flow influencing
the dynamics of all the other particles.

In the simulations, the friction coefficient $\gamma$ is taken to be equal to $2m/\tau$
where $\tau =\sqrt{m \sigma^2 / \epsilon} \approx 3 ps$ is the characteristic time scale of
oscillations in the Lennard-Jones well. The parameter $\sigma$=5 {\AA} used in the above
definition is a characteristic value of $\sigma _{ij}$ in the system. The selected value of
$\gamma$ corresponds to a
situation in which the inertial effects are small \cite{veit,Hoang} but the damping action
is
not
yet as strong as in water.
The equations of motion are solved by a fifth order predictor-corrector scheme.

Let us note, however, that although Langevin dynamics is commonly used in simulations of
biological systems, the validity of this approach is not always well-established. Namely,
as already noted by Lorentz \cite{Lorentz}, the Langevin equation in the above form may
only be used if there is a separation of time scales between the relaxation time of
the  particle (i.e. the bead)
velocity $\tau_v=\frac{m}{\gamma}=\frac{2a^2 \rho}{9 \eta}$ and the viscous relaxation
time of the solvent, $\tau_{\eta}=a^2 \frac{\rho_s}{\eta}$, where $\rho$ is the density of
the particle, $a$ its radius, and $\rho_s$ - the density of the solvent (see also the
thorough discussion in the book by Mazo \cite{Mazo}.)
The ratio of those two time scales is proportional to $\rho/\rho_s$. Since the  densities
of
proteins are only about 50\% higher than those of the surrounding liquid \cite{Tsai}, there
is no separation of time scales between the relaxation of fluid variables and those of the
bead and, strictly speaking, instead of
Eq.(\ref{lang}) one should use the generalized Langevin Equation involving a memory kernel
\begin{equation}
m \ddot{{\bf r}_i}(t) = -\int_{0}^t dt' \xi(t-t') (\dot{{\bf r}}_i(t')-{\bf u}({\bf
r}_i,t')) + {\bf F}^c_i(t) + {\bf \Gamma}_i(t) \;\;.
\end{equation}
where the noise is again Gaussian and related to the
dissipative term through the generalized fluctuation-dissipation relation
\begin{displaymath}
\langle {\bf \Gamma}_i(t) {\bf \Gamma}_j(t^\prime) \rangle =  k_BT \xi
(t-t^\prime) {\bf I} \delta_{ij}.
\end{displaymath}
Such an approach is naturally much harder to implement (see, however, \cite{Min})
 and thus the ordinary Langevin description as in Eq.(\ref{lang}) is
usually resorted to. In this paper, we show that in protein unfolding simulations the
results
of a simple
Langevin Dynamics (\ref{lang}) are consistent with those of Brownian Dynamics (see Sec. 4).
 The latter is not affected by the solvent and particle inertia effects, hence the
agreement
between the two methods seems to imply that
non-instantaneous response of the solvent to the change of particle velocity does not
play any important role in the protein unfolding processes.

\section{Brownian dynamics (BD)}

If the momentum relaxation time scale ($\tau_v$) is small in comparison to
the time scales characterizing the conformational evolution of the system ($\tau_c$),
it is appropriate to describe the dynamics
in terms of equilibration of the
particle configurations only. The exact definition of $\tau_c$ is problem dependent: if the
protein is stretched by the flow $U$ then $\tau_c=\frac{a}{U}$, if
a force $F$ acts on the molecule then $\tau_c=\frac{a \gamma}{F}$, and if the Brownian
diffusion
plays the central role in particle evolution then $\tau_c=a^2/D$, (where $D$ is the
diffusion
constant and $a$ the radius of the bead). The algorithm for simulations
of the evolution of particle positions in this time regime has been devised
by Ermak and McCammon \cite{ermak}. The displacement of particle $i$ in
time step $\Delta t$
(in the absence of the flow) obeys
 \begin{equation}
 {\bf r}_i  - {\bf r}_i^0 =  \sum_j \bigl( \nabla_j \cdot {\bf D}_{ij}^0 \bigr) \Delta t
+ \frac{1}{k_B T} \sum_j {\bf D}_{ij}^o \cdot {\bf F}^{0}_j \Delta t + {\bf B}_i,
\label{nar1}
\end{equation}
where the index $0$ denotes the values of respective quantities at the beginning of the
time step, ${\bf F}_j$ is the force exerted on particle $j$ by
other particles,
${\bf D}$ is the diffusion tensor and  ${\bf B}$ - a random displacement given
by a Gaussian distribution with an average value of zero and covariance
obeying
 \begin{equation}
<{\bf B}_i {\bf B}_j> = 2 {\bf D}^0_{ij}
\Delta t.
\label{Gauss}
\end{equation}

It is nontrivial to generalize the above expression to incorporate the effects of a
general external flow field \cite{ansell}. However, in the present case, we will be
interested only
in
the uniform
flow, in which case one gets simply
 \begin{equation}
 {\bf r}_i - {\bf r}_i^o = {\bf U} \Delta t  + \sum_j \bigl( \nabla_j \cdot {\bf D}_{ij}^0
\bigr) \Delta t
+ \frac{1}{k_B T} \sum_j {\bf D}_{ij}^o \cdot {\bf F}_j^0 \Delta t + {\bf B}_i,
\label{nar}
\end{equation}
where ${\bf U}$ is the flow velocity.
If the diffusion tensor is nondiagonal, there exists a coupling between the force acting on
the particle $j$ and the displacement of particle $i$ ({\it cf.} Eq.\ref{nar1}). This
coupling, mediated by the solvent, is commonly
referred to as the ``hydrodynamic interactions''.

Note that without the HI, the diffusion tensor is simply
\begin{equation}
{\bf D}_{ij}=\frac{k_B T}{\gamma} {\bf I} \delta_{ij}
\label{nohi}
\end{equation}
and we recover the overdamped limit of Eq.(1).
The diffusion tensor ${\bf D}$ depends in a complicated nonlinear way on the instantaneous
positions of all particles in the system. For a system of spheres, exact explicit
expressions for the diffusion tensor ${\bf D}_{ij}$ exist in the form of the power series
in
interparticle distances, which may be incorporated into the simulation scheme
 \cite{
Kim,Mazur,Brady,Felderhof,Ladd,Cichocki,Brady2}. Here we adopt a pairwise, far-field
approximation
of
${\bf D}$ proposed by Rotne, Prager \cite{Rotne} and Yamakawa \cite{Yamakawa}

\begin{equation}
{\bf D}_{ii}=\frac{k_B T}{\gamma} {\bf I}
\label{rp0}
\end{equation}
and
\begin{equation}
{\bf D}_{ij}=\displaystyle \frac{k_B T}{\gamma} \frac{3a}{4 r_{ij}}
\left\{\begin{array}{cl}
      \displaystyle \left[ \left( 1+\frac{2a^2}{3 r_{ij}^2} \right) {\bf I} + \left(1-
\frac
{2a^2}{r_{ij}^2}\right)
{\bf \hat{r}}_{ij} {\bf \hat{r}}_{ij} \right], & r_{ij} \geq 2a\ \\
\\
\displaystyle \frac{r_{ij}}{2a} \left[ \left( \frac{8}{3}-\frac{3r_{ij}}{4a} \right) {\bf
I}
+
\frac
{r_{ij}}{4a}
{\bf \hat{r}}_{ij} {\bf \hat{r}}_{ij} \right], & r_{ij}  < 2a
         \end{array}\right.
\label{rp}
\end{equation}
where ${\bf r}_{ij}={\bf r}_{j}-{\bf r}_{i}$  and $a$
represents
the hydrodynamic radius of a bead. Since the above expression is exact only in the large
$r_{ij}$ limit,
the radius $a$ should be taken to be significantly smaller than 1.9 \AA, which is the half
of
the distance
between the succesive beads. On the other hand, $a$ cannot be too small, since the space
along the chain
is densely filled with amino acids. We take $a=1.5$ {\AA} in our simulations, which seems a
reasonable
starting point for a qualitative assessment of HI impact on protein unfolding. However,
further studies
on the impact of $a$ on the system dynamics are needed, in particular the hydrodynamic
radius
might
need to vary along the chain, reflecting the different sizes of the residues.

In the approximation (\ref{rp}), the divergence of
the diffusion
matrix vanishes ($\nabla_j \cdot {\bf D}_{ij} \equiv 0$), which further simplifies the
numerical
scheme. However, if the full hydrodynamic interactions are included, the divergence term
should be taken into account \cite{Wajnryb}.

The simulation using Eq. (\ref{nar1}) together with  (\ref{rp0}) and  (\ref{rp}) will be
referred to as Brownian Dynamics with hydrodynamic interactions (BDHI) in contrast to a
simple BD with the
diagonal diffusion tensor (\ref{nohi}).

Note that the BD describes configurational evolution of the beads on time scales in which
the
inertia effects of the beads and solvent molecules are negligible \cite{Naegele} and,
therefore, time scale separation issues discussed in Section 3 are not pertinent here.
This feature favours BD as a
method of choice when simulating stochastically driven motion of proteins
at a coarse-grained level \cite{Dickinson}.

In our previous studies on ubiqutin unfolding \cite{cs1,cs2}, we have used LD.
Here, on the other hand, we incorporate HI within the BD approach.
This calls for a comparison of the three schemes (LD, BD, and BDHI) to
distinguish the effects resulting from HI and those from the usage of distinct
integration schemes.

\section{Constant velocity stretching}

Fig. 1 presents the force-extension curves for the constant-velocity unfolding at
different unfolding speeds.
In the simulations, both termini of a protein are attached to harmonic springs
with the elastic constant $k$=0.06 $\epsilon /${\AA}$^2$. The other end of the N-terminus
spring is fixed whereas the
C-terminus spring moves at a speed $v_p$.
We consider three values of $v_p$: 0.5 , 0.05  and
$v_p=0.005 $\AA$/\tau$.

In the low-speed limit, all the three data sets obtained using the LD, BD and
BDHI are seen to converge to a single curve. In contrast, at large unfolding speeds,
the differences between the LD and BD are pronounced. However, strictly speaking, in this
time
regime BD has a limited validity, because of the lack
of separation between the momentum relaxation time ($t_v=\frac{m}{\gamma}=0.5 \tau$) and
the characteristic time of the aminoacid movement due to the stretching
$t_{pull}=\frac{a}{v_p} = 3 \tau$ (for the highest speed quoted above).
In the experiments, the separation of time scales is huge. In water
$\gamma \approx 6 \pi \eta a \approx 3 \times 10^{-9}$ g s$^{-1}$ which leads to
$\tau_v = \frac{m}{\gamma} \approx  0.06$ ps (for the typical aminoacid mass of
$m \approx 2 \times 10^{-22}$ g). On the other hand, the pulling speeds are of the order of
500 nm/s which gives $\tau_p \approx 0.3$ ms, thus there is a five-order-of-magnitude
separation in time scales.
For such a case, LD and BD simulations would give exactly the same result.

The fact that the differences between the BD and BDHI trajectories disappear
in the limit of small $v_p$ is due to the lack of impact of HI at small velocities.
To conclude,
in the experimentally relevant small speed limit, the effects
of HI are expected to be negligible.

An inspection of Fig. 1 indicates that
in the case of high stretching speeds neglecting HI results in larger peak forces. In
the high speed all-atom simulations of titin in water \cite{Schulten}
the forces of stretching are found to be excessively large.
Such all-atom molecular dynamics programs are not geared towards
hydrodynamic phenomena and may incorporate HI poorly.
It is possible to consider that the excessive force could be partially due to
the missing HI.

\section{Force clamp unfolding}

In the force-clamp AFM unfolding \cite{FernandezLi,Schlierf} one
applies the stretching force to the protein terminus and monitors the
end-to-end distance, $L$.
The experimental data and the numerical simulations \cite{cs1,West} show that
proteins unfold in a stepwise manner
at a constant force. This means that a rapid
unfolding transition takes place after a certain waiting time.
The smaller the force, the longer the waiting time.

In our simulations, we apply the force to the $C$ terminus of the protein,
whereas the N-terminus is attached
to a harmonic spring of elastic constant $k$=0.06 $\epsilon /${\AA}$^2$.
The unfolding trajectories of ubiquitin are presented in Fig. 2
for two values of the force,
$F=2.4\ \epsilon/${\AA} and $F=4\ \epsilon/${\AA}.
The LD and BD methods essentially coincide
for these relatively large forces
(small differences between the trajectories are merely stochastic in nature).
However, the inclusion of HI changes the physics considerably -- the waiting times
become much smaller and the duration of the unfolding transition itself decreases from
$\approx 250 \tau$ to about $50 \tau$ at both values of the force.

The fact that the HI facilitate protein unfolding
may be understood qualitatively when one realizes that
an amino acid moving away from the bulk of a protein creates a flow which drags other
residues with it ( see Fig. 3).

The differences between unfolding with and without HI  are further highlighted by
analysis of the so called unfolding scenarios \cite{Hoang}, in which one plots
an average time when a given contact is broken against the contact
order, i.e. against the sequential distance, $|j - i|$, between the
amino acids that form a native contact. Figs. 4 and 5 compare the unfolding scenarios
for LD, BD and BDHI at $F=2.4\ \epsilon/${\AA} and $F=4\ \epsilon/${\AA} respectively.
Remarkably, although the differences in time scales between the unfolding with and without
HI are considerable, the unfolding scenarios for the smaller force
are similar (Fig. 4), which shows that the unfolding pathway of a protein is not affected
by
the hydrodynamic effects. However, as the force is increased, both LD and BD scenarios
change
(Fig. 5): the $\beta$ hairpin structure now unfolds at the end instead of at the beginning
of
the unfolding process. In contrast, in the case of BDHI, such a switch is not observed:
the scenarios for larger and smaller forces look qualitatively similar.

\section{Unfolding in a uniform flow}

Finally, we study the influence of HI on the characteristics of the
protein unfolding in a uniform flow with a speed of $U$.
Although the process has not yet been
realized experimentally,
the simulations \cite{lemak1, lemak2,cs2} seem to suggest that uniform flow unfolding
leads to a richer set of metastable
conformations than the constant force pulling. For example,
when the N terminus is anchored, ubiquitin stretches in a flow
through two distinct intermediate
states corresponding to a partial unzipping.

A detailed analysis of the uniform flow unfolding of ubiqutin in the absence of
HI, together with the snapshots of intermediate
conformations for different anchorings, can be found in \cite{cs2}.
In that reference, we have mistakenly reported values of the forces in wrong
units. Instead of the correct unit of $\epsilon$/\AA\ we used $\epsilon /\sigma$,
where $\sigma $ was equal to 5\AA.

Fig. 6 shows examples of unfolding trajectories of ubiquitin in a uniform flow for four
different flow velocities, both with and without HI.
We observe that unfolding of
the system with HI requires a much larger flow speed than without.
This can be understood qualitatively
in terms of the so-called no-draining effect \cite{Rzehak}: the residues hidden inside the
protein
are shielded from the flow and thus only a small fraction of the residues experience the
full
drag force of $F=-\gamma U$ (see Fig. 7).
In contrast, when no HI are present, this drag force is applied to all
residues \cite{cs2}.

Notably, although the time scales and velocities involved in the protein unfolding with
and without HI are completely different, the metastable states are nearly
identical (see Fig. 6).
This feature is related to the dynamic character of HI - they  do not change the
potential energy of the system and, therefore, do not affect its stationary properties.
In principle, however, one could imagine a situation in which, due to the differences in
dynamics imposed by HI, a system chooses alternative pathways when unfolding.
This is clearly not the case here which may be related to the {\em directed}
character of the disturbance (flow in this case or the force in force-clamp
experiment) which imposes a prefered direction of unfolding, thus greatly reducing the set
of available unfolding pathways.

The similarities in unfolding pathways of ubiquitin
are further confirmed by the comparison of unfolding
scenarios. As an example, Fig. 8 shows the unfolding scenario
for the flow $U=3.5$ \AA$ /\tau$  with HI
compared to $U=0.55$ \AA$ /\tau$ without HI (the mean unfolding times
are comparable in both cases).
The scenarios are very close to each other, the main difference being that
the HI enhance cooperativity by breaking the contacts in a more simultaneous fashion.

\section{Summary}

In summary, hydrodynamic interactions seem to affect the time scales of unfolding by a
constant force
and by a fluid flow in opposite way but keep the set of the possible
metastable states. The HI may also reduce peak forces in stretching
at a constant speed, although this effect weakens with the diminishing stretching speed.

This work has been supported by the European program IP NaPa through Warsaw
University of Technology.

%\begin{references}

\newpage
\centerline{FIGURE CAPTIONS}

\begin{description}

\item[Fig. 1. ]
Force-extension curves for the ubiquitin unfolding at a
constant speed obtained using Langevin Dynamics (dotted line) and Brownian Dynamics
with (thick solid line) and without (thin solid line) hydrodynamic interactions.
The succesive panels correspond to the
pulling speeds of $v_p=0.5, 0.05$ and $0.005\ $\AA$/\tau$
from top to bottom respectively.

\item[Fig. 2. ]
The end-to-end distance of the model ubiquitin as a
function of time during unfolding at
a constant force. The thick solid, thin solid and dotted lines
correspond to the BDHI, BD, and LD simulations respectively.
The upper panel corresponds to the force $F=2.4\ \epsilon/${\AA}, and the
lower one to $F=4\ \epsilon/${\AA}.

\item[Fig. 3. ]
The dragging effect: the moving particle creates a flow patern which
affects other particles
by pulling them in the direction of its motion.

\item[Fig. 4. ]

Unfolding scenarios of ubiquitin at constant force of  $F=2.4 \ \epsilon/${\AA}
simulated with LD (the upper panel),
BD (the center panel) and BDHI (the lower panel).
Open circles, triangles, squares,  pentagons
and solid triangles and squares correspond to contacts (36-44)--(65-72),
(12-17)--(23-34), [(1-7),(12-17)]--(65,72), (41-49)--(41-49),
(17-27)--(51-59), (1-7)--(12-17) respectively.
The crosses denote all other contacts.
The segment (23-34) corresponds to a helix. The two $\beta$-strands
(1-7) and ((12-17) form a hairpin. The remaining $\beta$-strands
are (17-27), (41-49), and (51-59).

\item[Fig. 5. ]

The same as in Fig. 3 but for $F=4\ \epsilon/${\AA}.

\item[Fig. 6. ]
The end-to-end distance of the model ubiquitin
as a function of time during unfolding
by a uniform fluid flow as illustrated by several trajectories.
The lower panel corresponds to the description with the
hydrodynamic interactions and the upper -- without.
The succesive curves in the upper panel correspond to the flows of
$0.25,0.3,0.4$ and $0.5 \ \text{\AA}/\tau$ bottom to top respectively.
In the lower panel the flow speeds are
$2.5, 3.5, 5.0, 10.0$ \AA$/\tau$
In all the simulations, the N-teminus of the protein is fixed.

\item[Fig. 7. ]
The shielding effect: the particles inside a cluster experience a
smaller drag force than
those on the surface.

\item[Fig. 8. ]

The left and right panels show unfolding scenarios without
the hydrodynamic interactions ($U=0.55$ \AA$/\tau$) and
with the hydrodynamic interactions ($U=3.5$ \AA$/\tau$)
respectively.

\end{description}

%FIGURE 1
\begin{figure}
\vspace{-4cm}
\includegraphics[width=12cm, viewport=30 100 500 400]{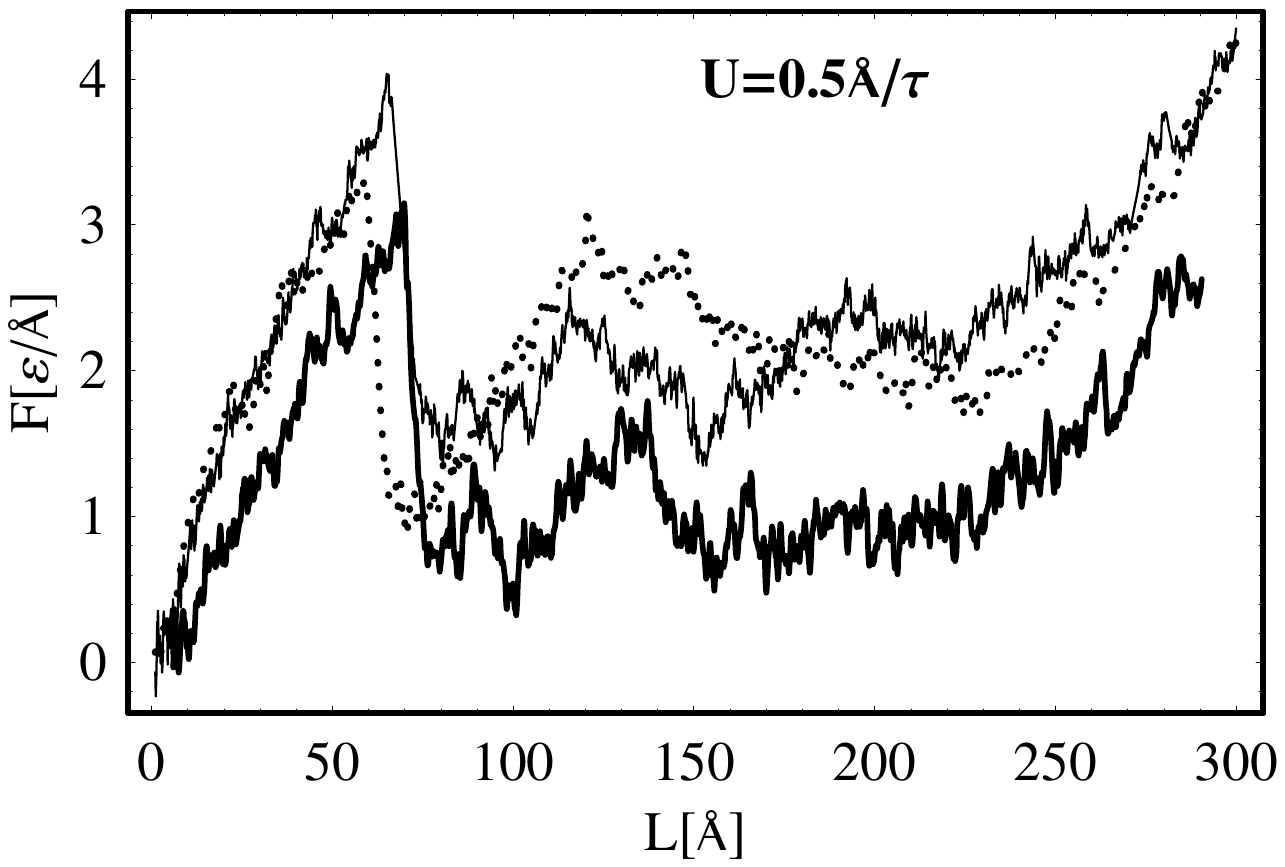}
\includegraphics[width=12cm, viewport=30 100 500 400]{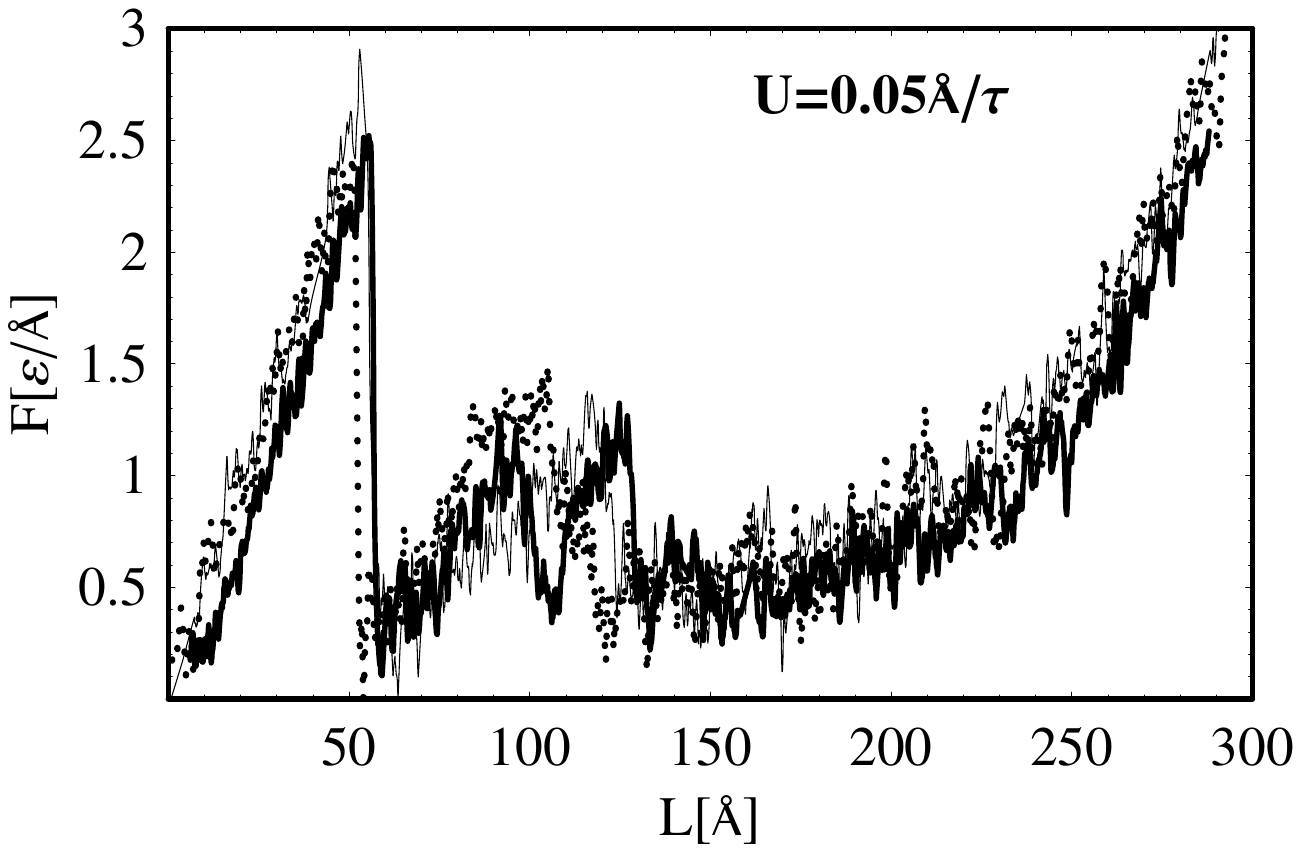}
\includegraphics[width=12cm, viewport=30 100 500 400]{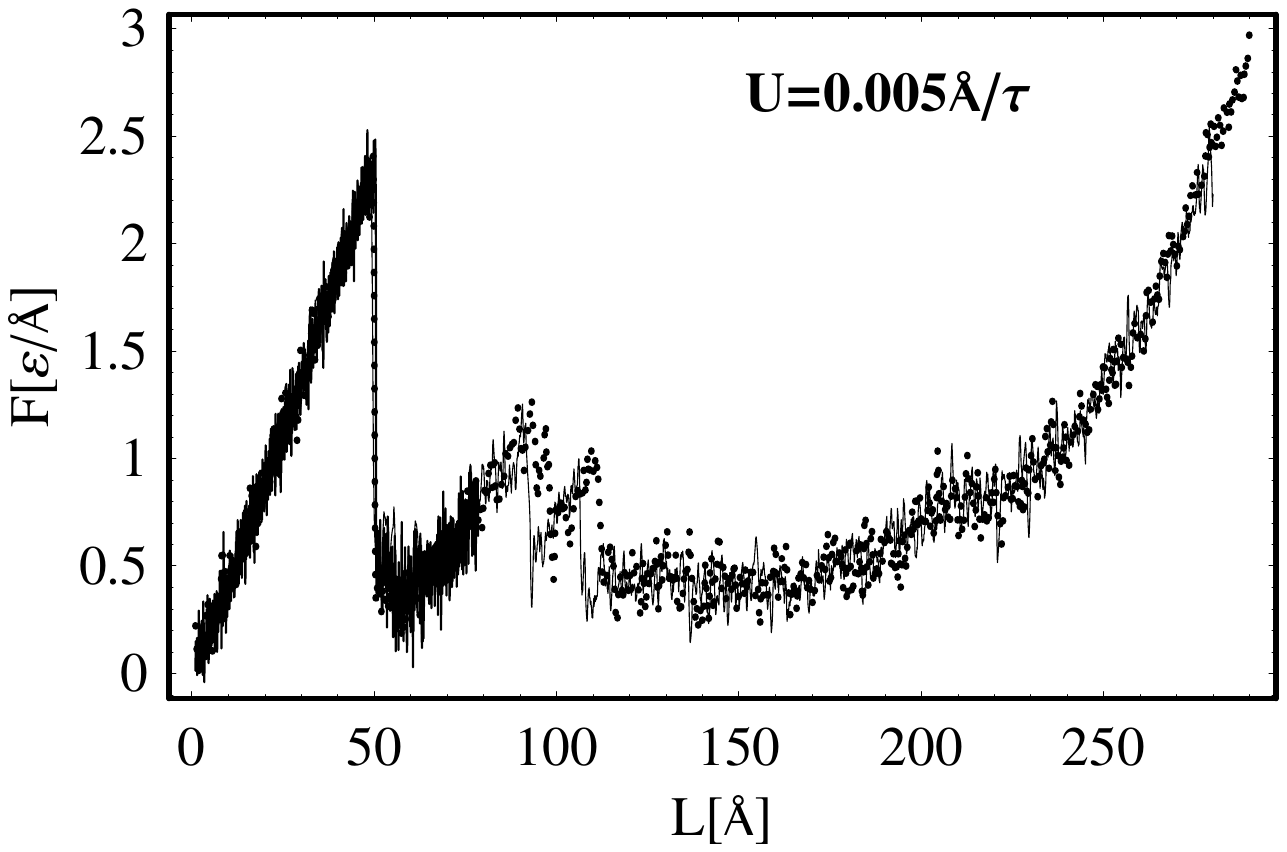}
\vspace*{3cm}
\caption{ }
\end{figure}

\vspace*{-4cm}

\clearpage

\newpage

%FIGURE 2
\begin{figure}
\includegraphics[width=12cm, viewport=30 100 500 400]{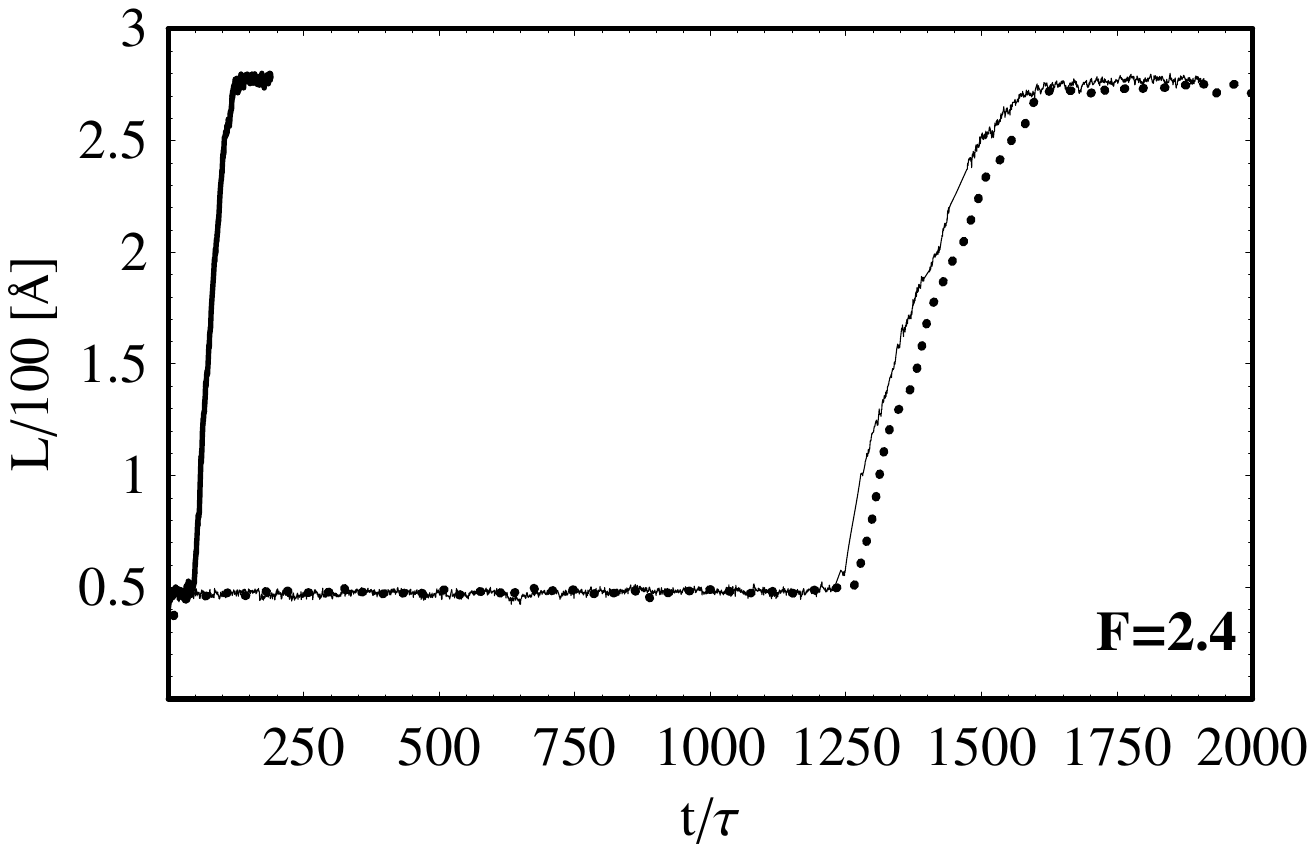}
\includegraphics[width=12cm, viewport=30 100 500 400]{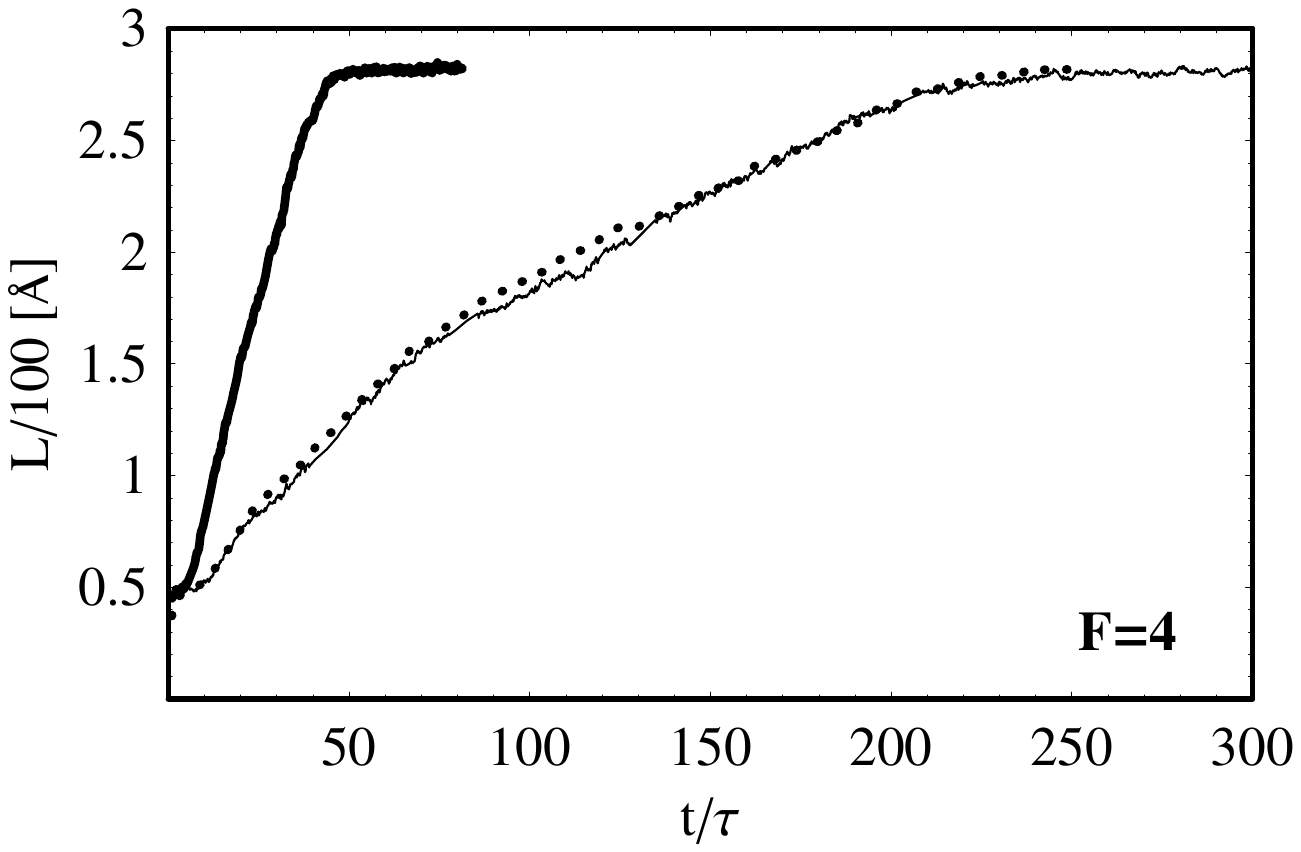}
\vspace*{3cm}
\caption{ }
\end{figure}

\vspace*{-4cm}

%FIGURE 3
\begin{figure}
\includegraphics[width=12cm, viewport=30 100 500 400]{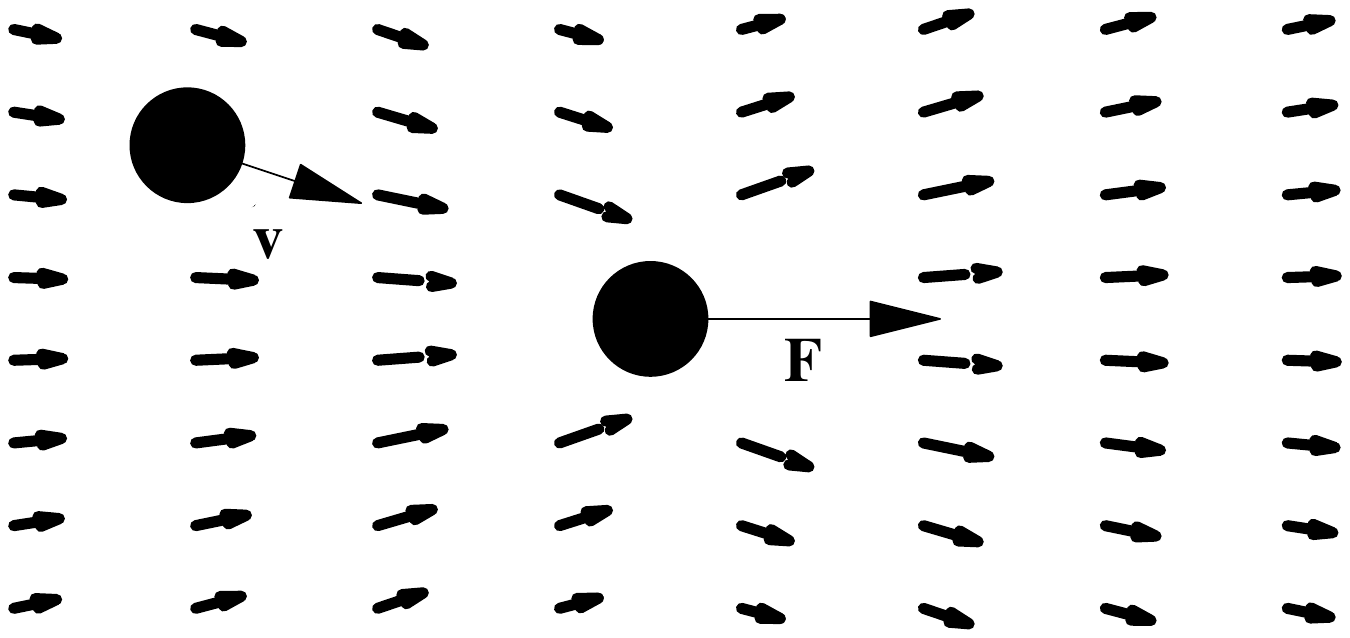}
\vspace*{3cm}
\caption{ }
\end{figure}

\vspace*{-4cm}

%FIGURE 4
\begin{figure}
\vspace{-5cm}
\includegraphics[width=12cm, viewport=30 100 500 400]{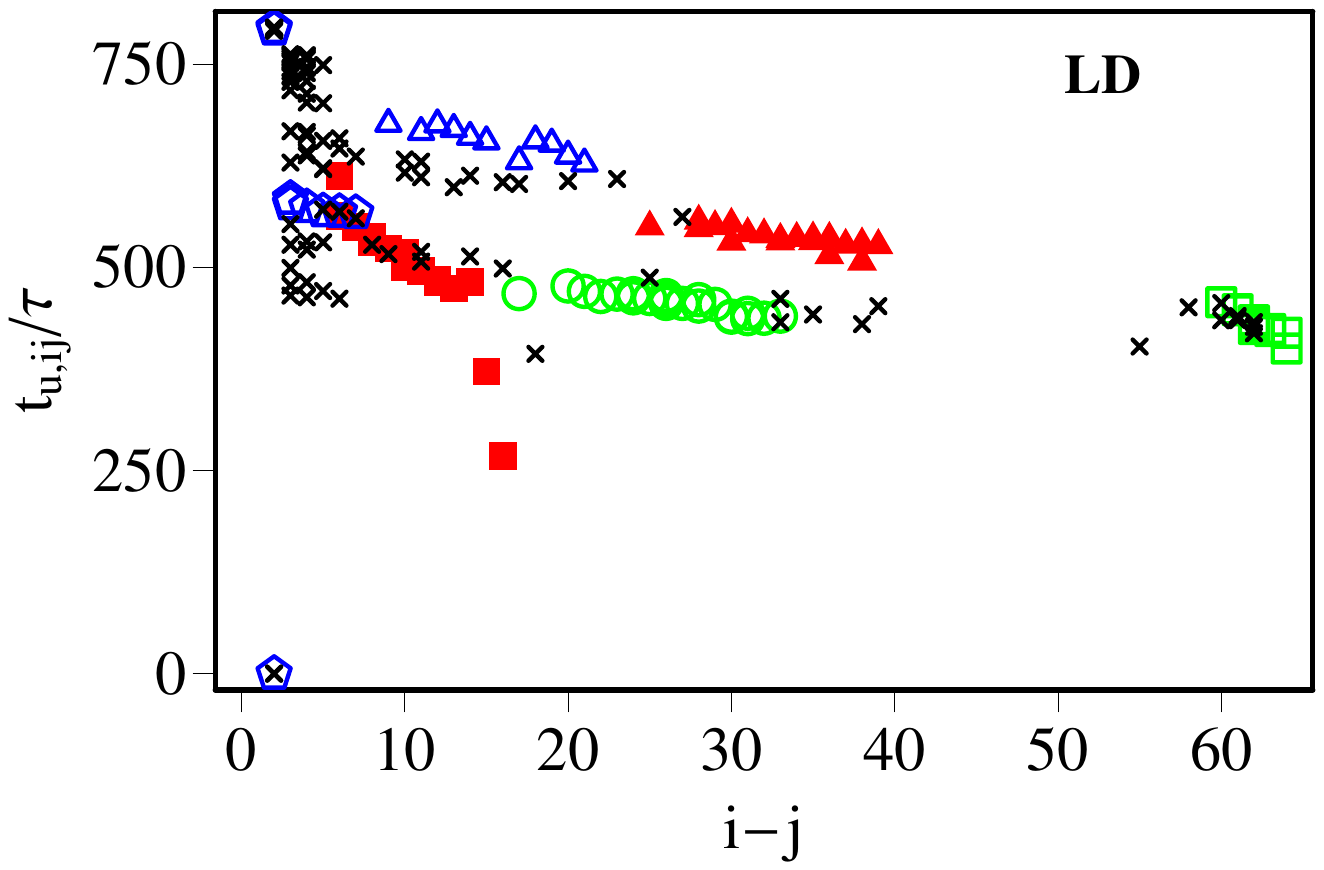}
\includegraphics[width=12cm, viewport=30 100 500 400]{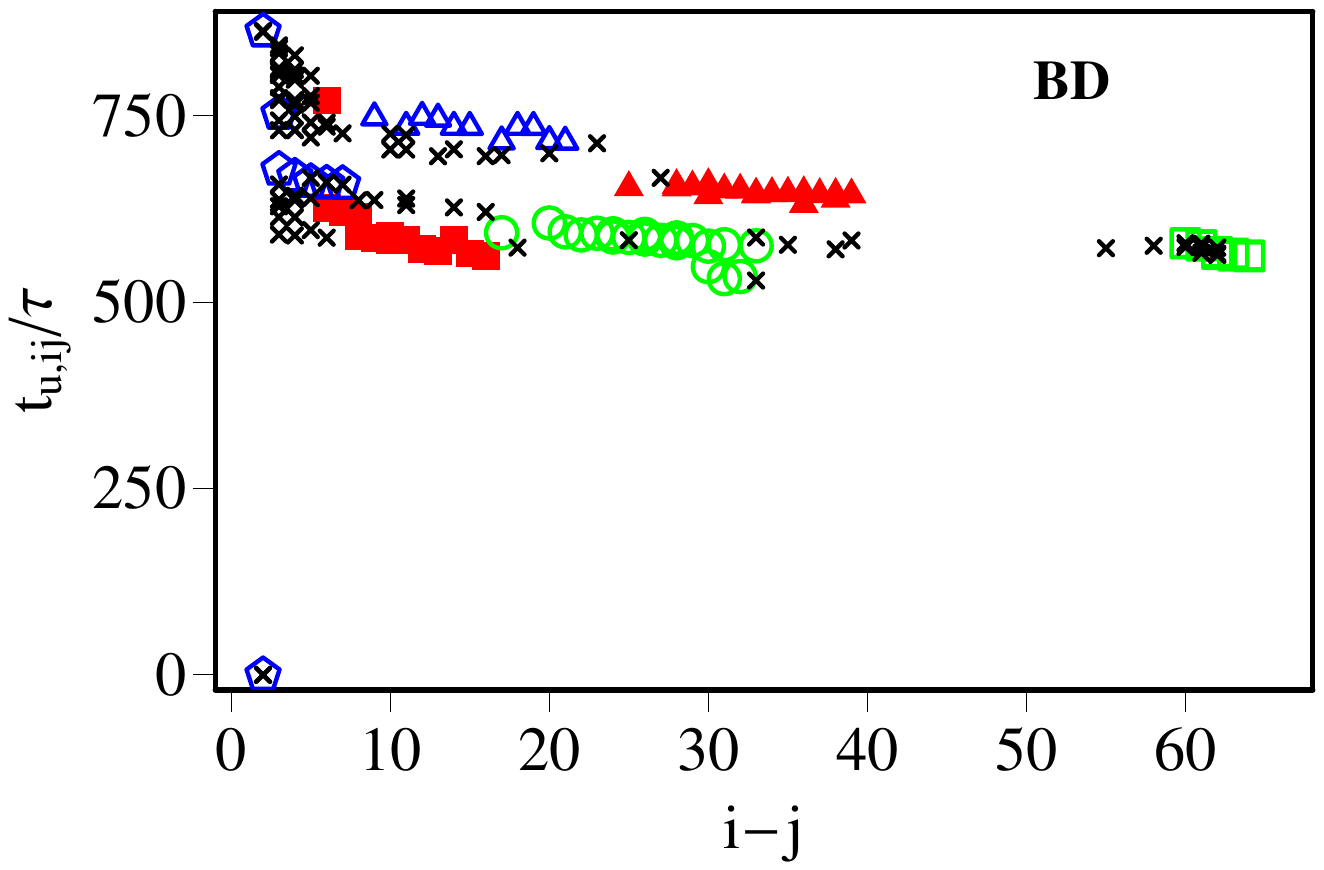}
\includegraphics[width=12cm, viewport=30 100 500 400]{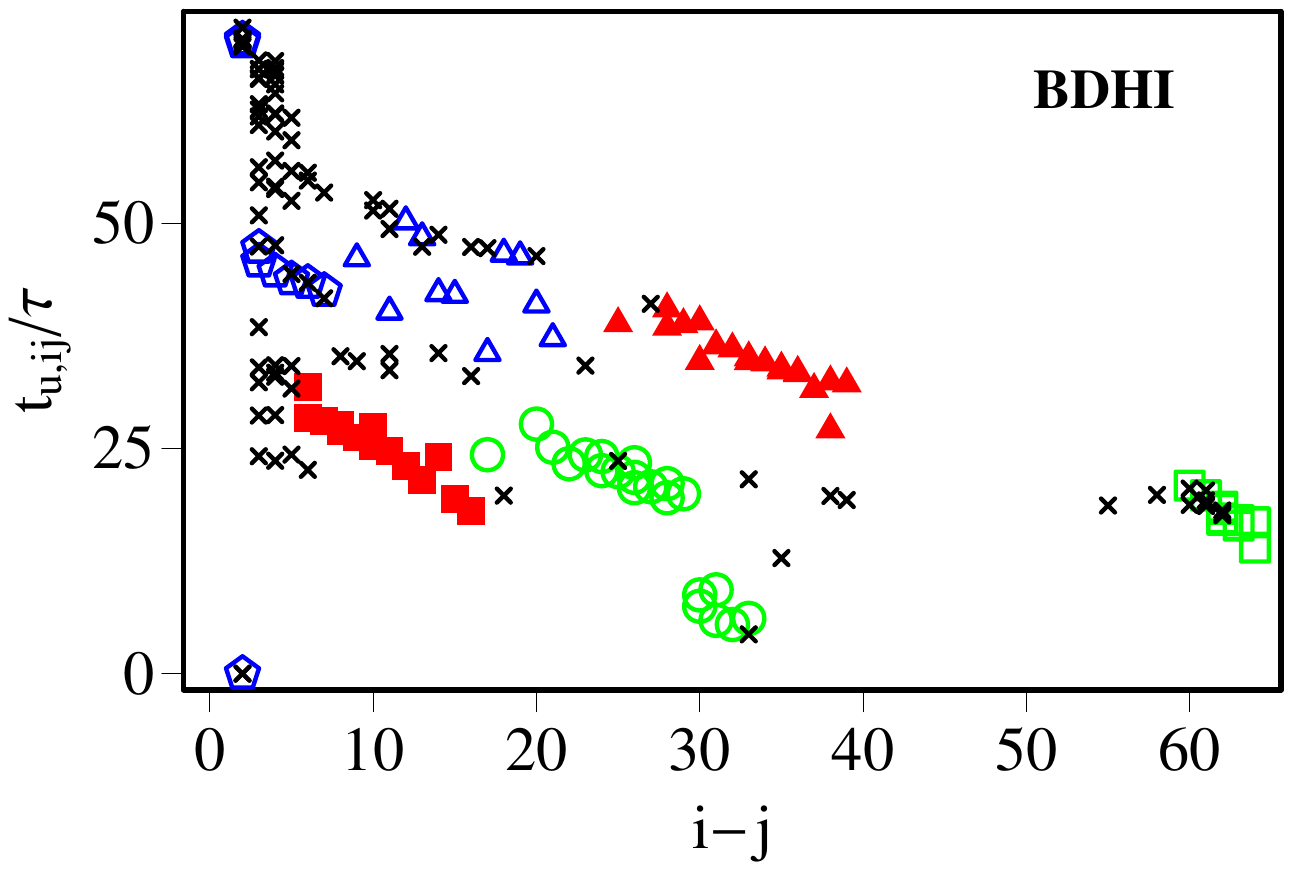}
\vspace*{3cm}
\caption{ }
\end{figure}

\vspace*{-4cm}

%FIGURE 5
\begin{figure}
\vspace{-5cm}
\includegraphics[width=12cm, viewport=30 100 500 400]{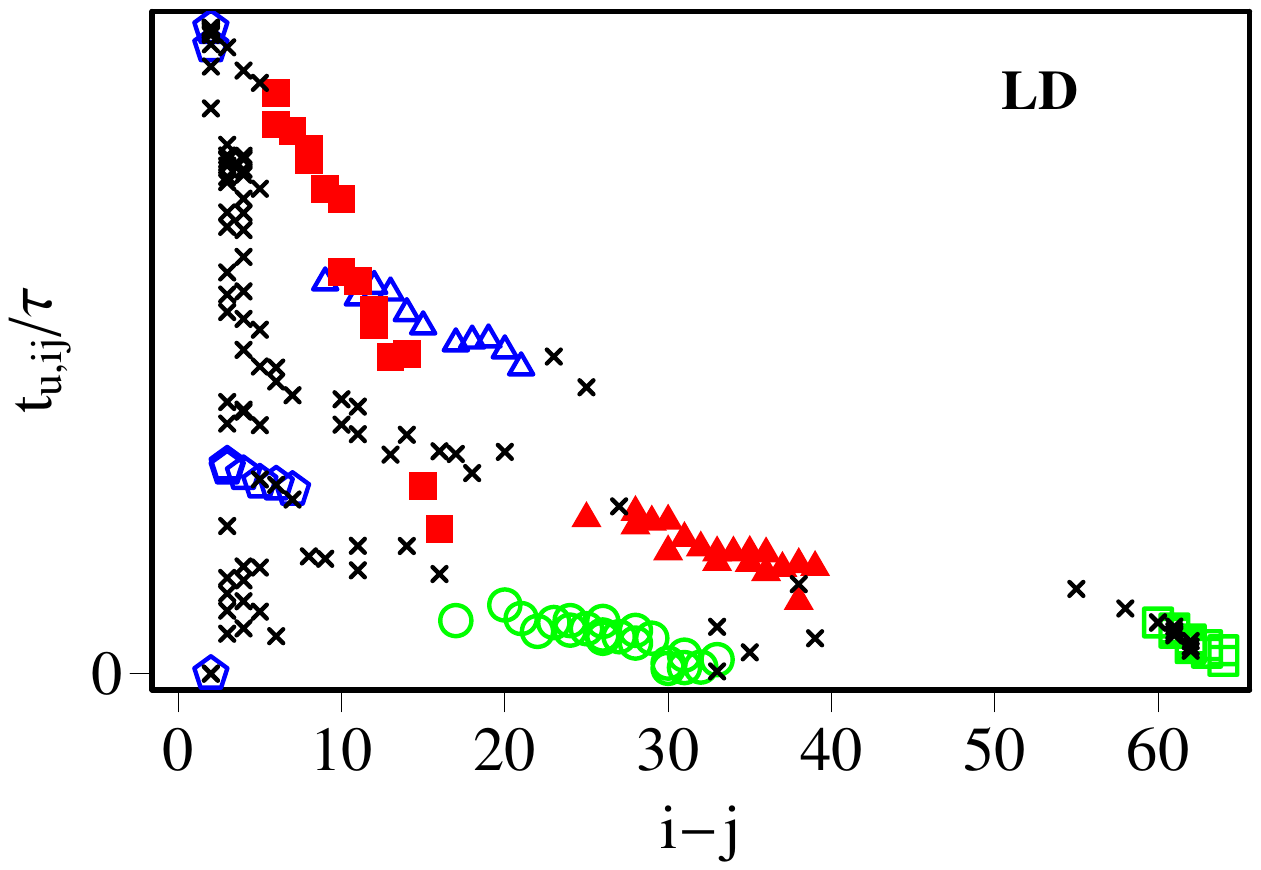}
\includegraphics[width=12cm, viewport=30 100 500 400]{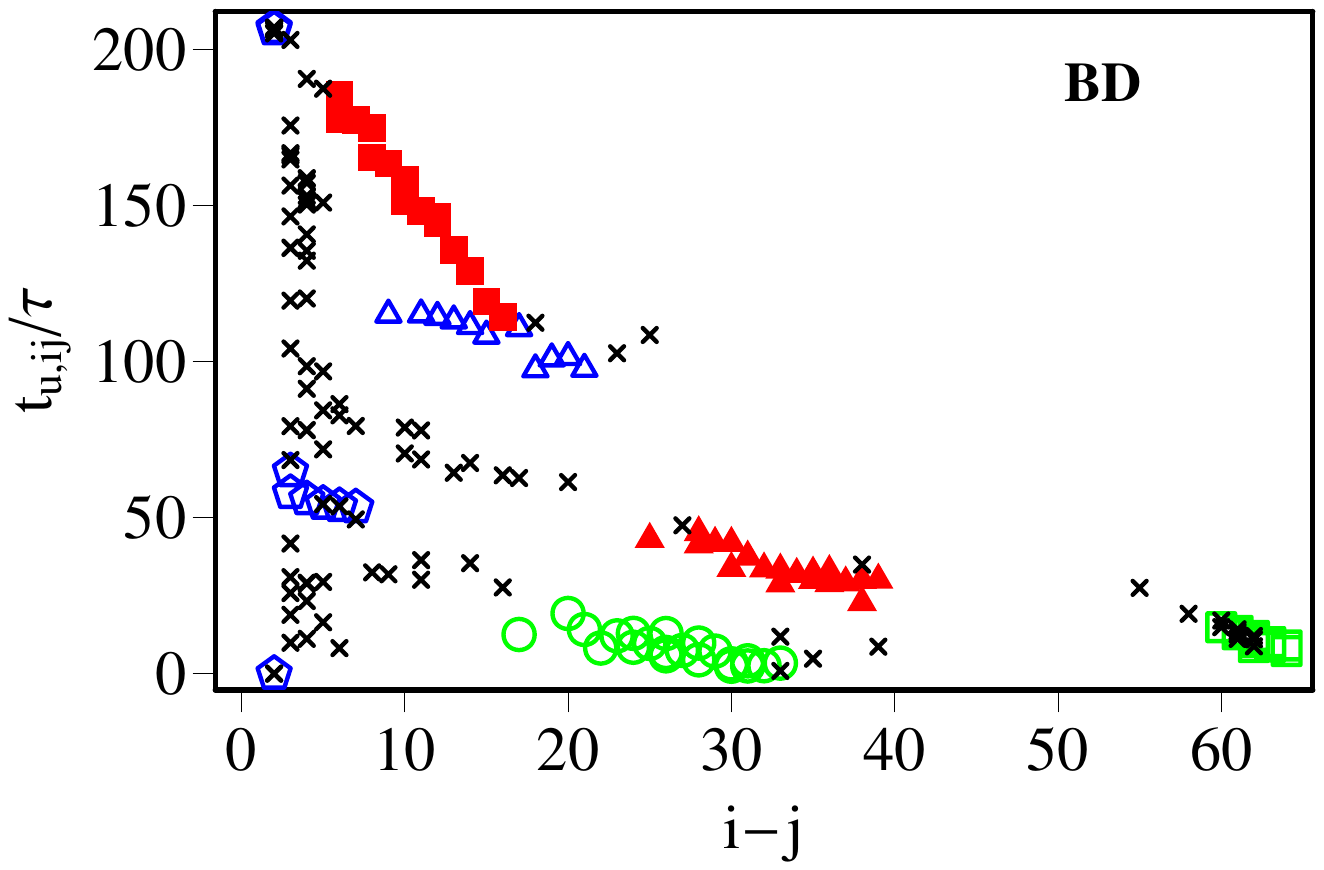}
\includegraphics[width=12cm, viewport=30 100 500 400]{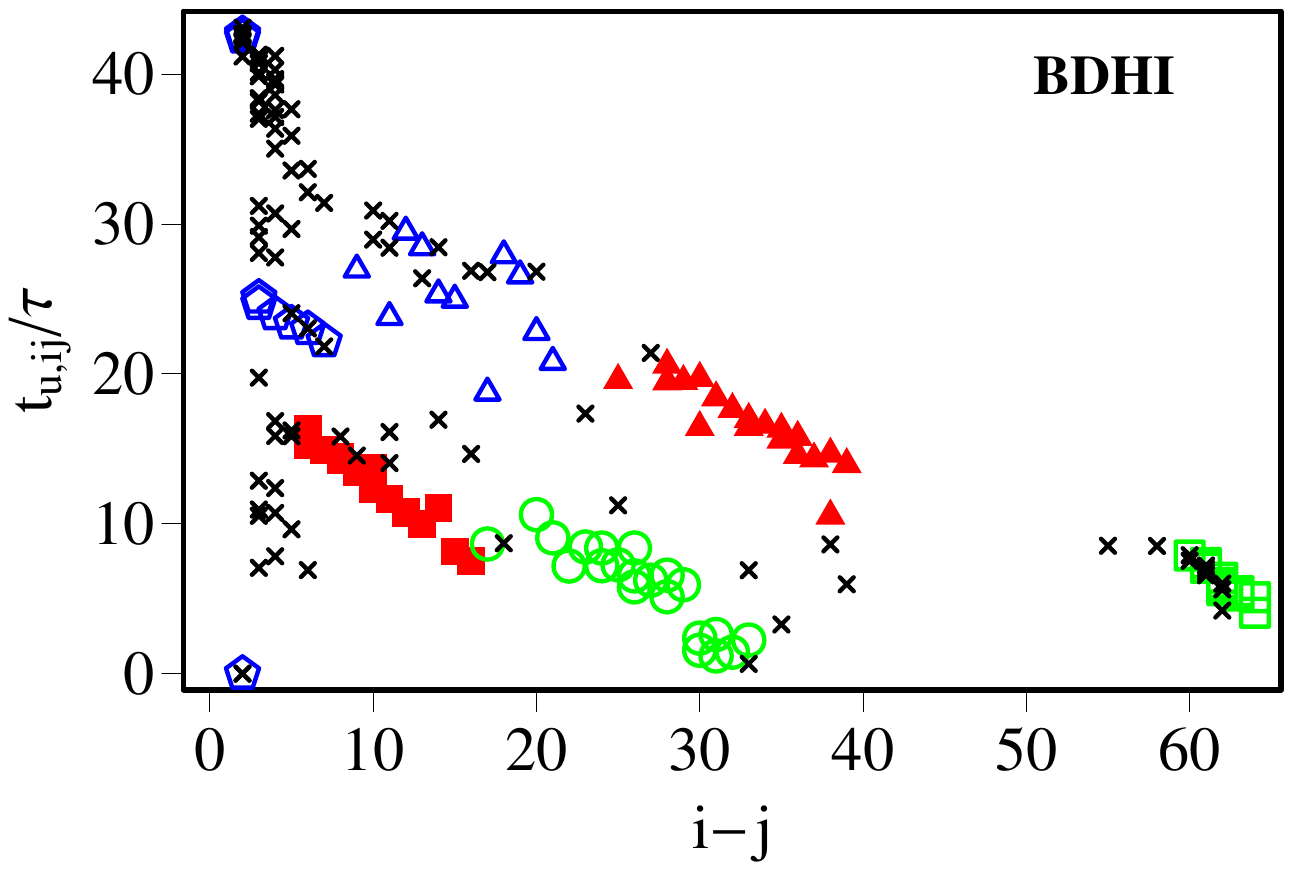}
\vspace*{3cm}
\caption{ }
\end{figure}

\vspace*{-4cm}

%FIGURE 6
\begin{figure}
\includegraphics[width=12cm, viewport=30 100 500 400]{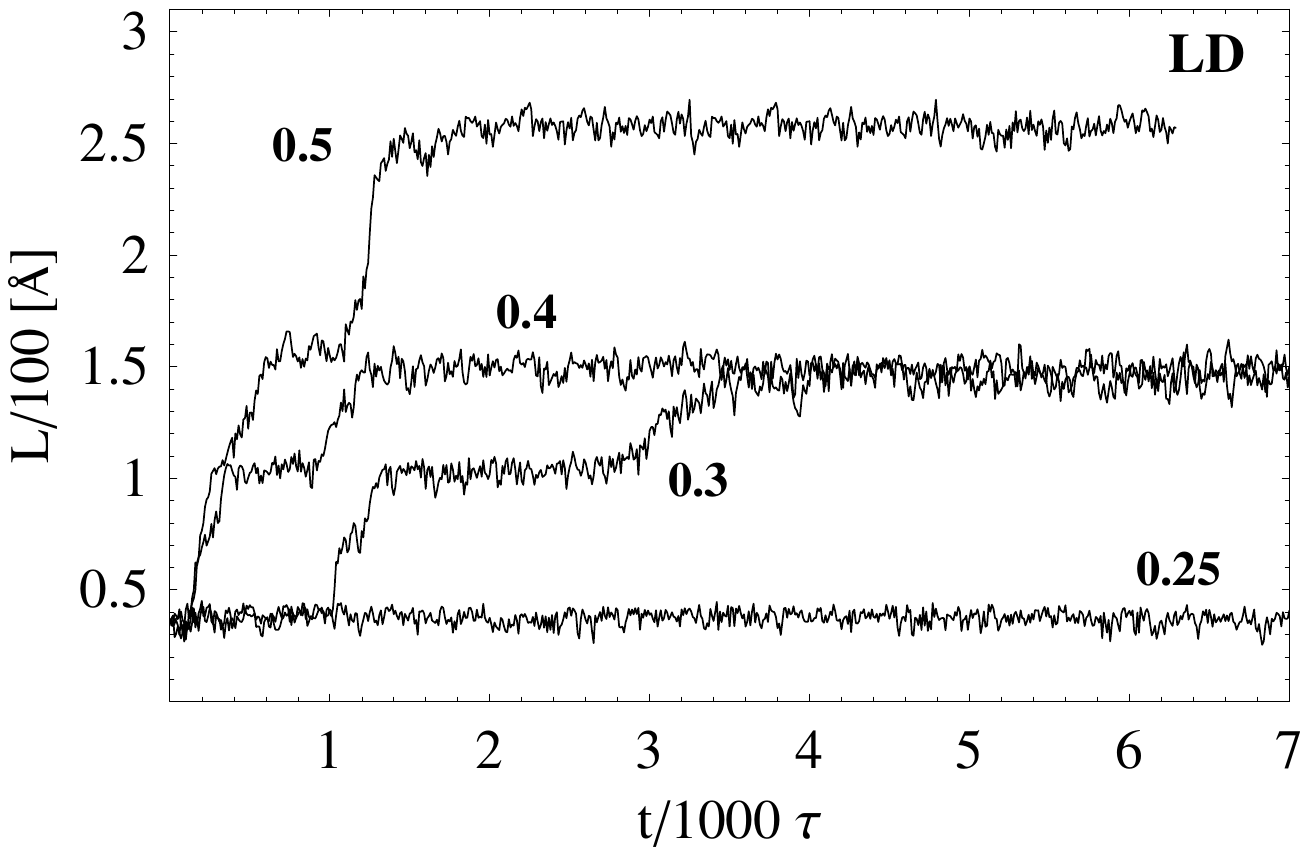}
\includegraphics[width=12cm, viewport=30 100 500 400]{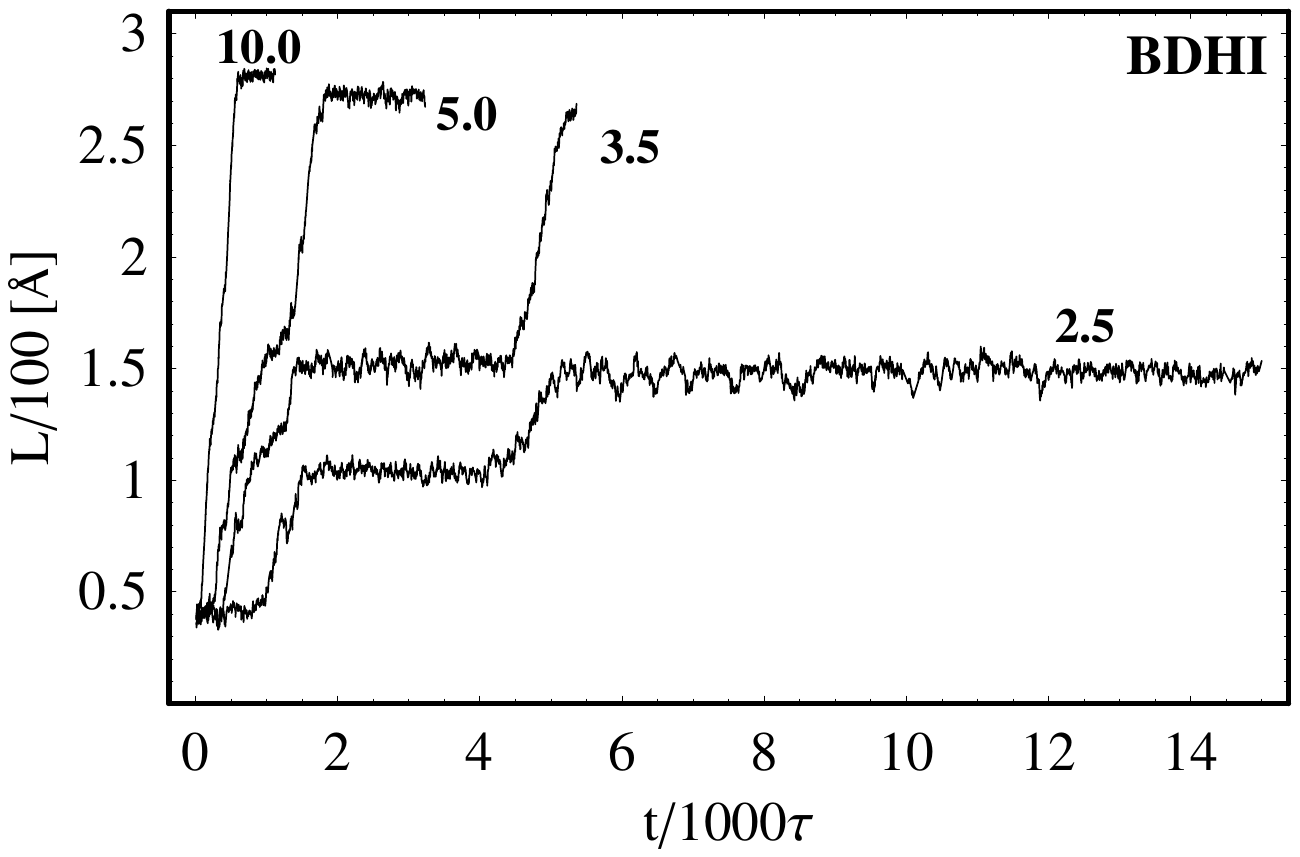}
\vspace*{3cm}
\caption{ }
\end{figure}

\vspace*{-4cm}

\clearpage

\newpage

%FIGURE 7
\begin{figure}

\includegraphics[width=12cm, viewport=30 100 500 400]{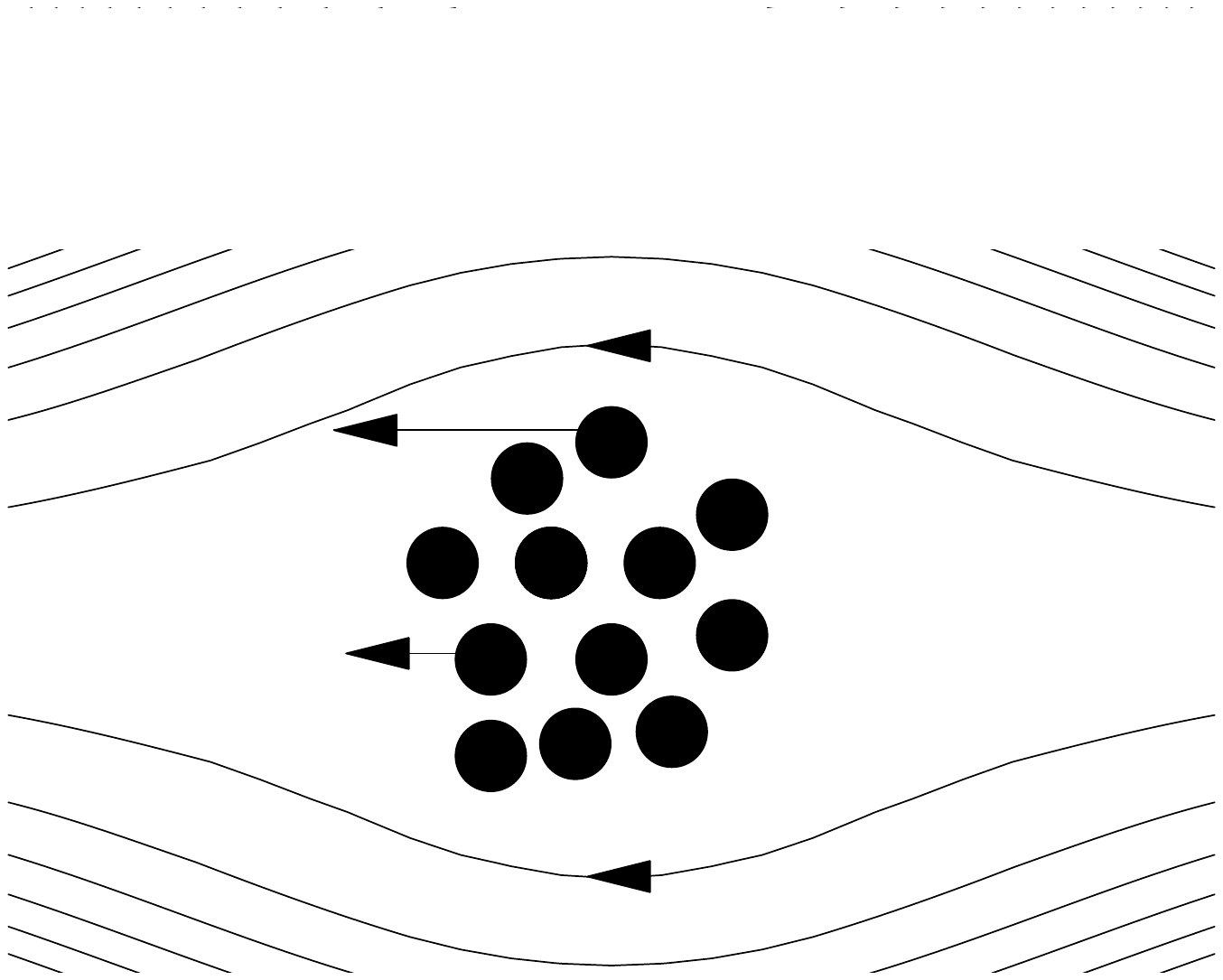}
\vspace*{3cm}
\caption{ }
\end{figure}

\vspace*{-4cm}

\newpage

%FIGURE 8
\begin{figure}
\includegraphics[width=15cm, viewport=100 100 600 400]{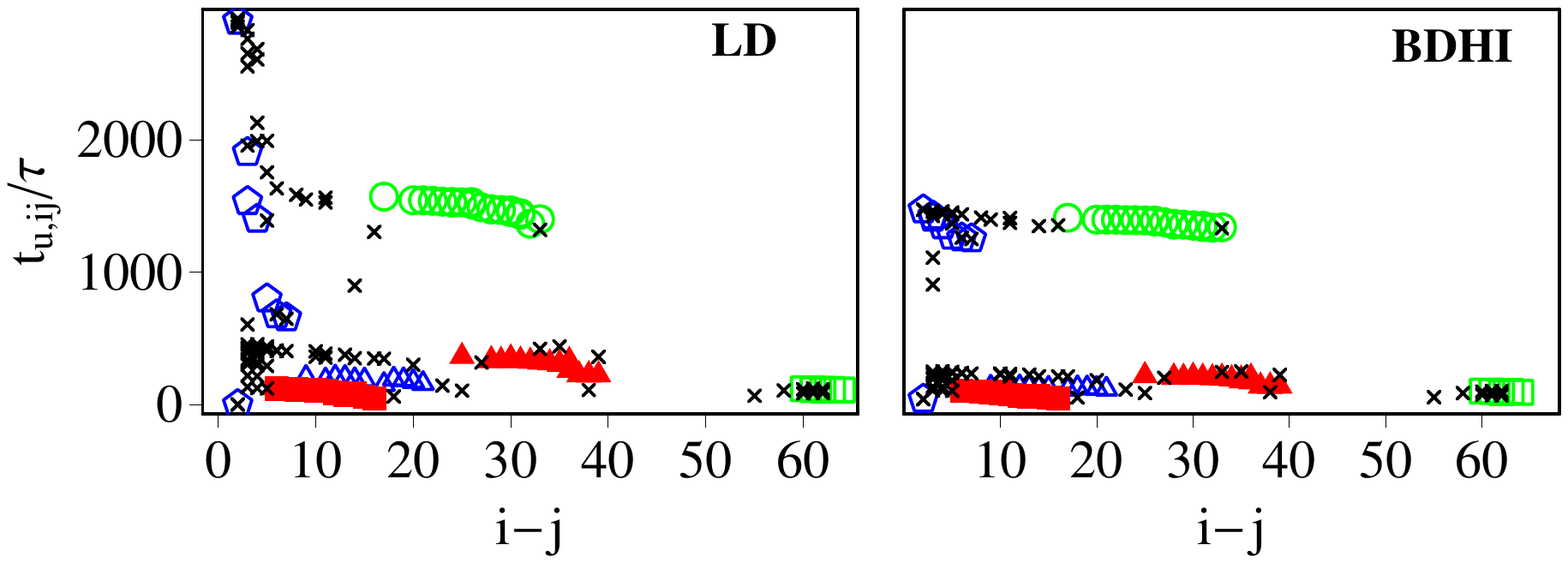}
\vspace*{3cm}
\caption{ }
\end{figure}

\vspace*{-4cm}

\clearpage

\newpage


\begin{thebibliography}{199}
%\begin{description}

\bibitem{Dhont}
J. K. G.  Dhont, An Introduction to Dynamics of Colloids (Elsevier, Amsterdam, 1996).

\bibitem{Larson}
R. G. Larson and J. J. Magda
{\it Macromolecules}, 22, 3004 (1989).

\bibitem{Tanaka}
H. Tanaka,
{\it J. Phys. Cond. Matt.}, {\bf 13}, 4637 (2001).

\bibitem{Dickinson}
E. Dickinson,
%Brownian dynamics with hydrodynamics interactions: the application
%to protein diffusional problems.
{\it Chem. Soc. Rev.}, {\bf 14}, 421, (1985).

\bibitem{Wojtaszczyk}
P. Wojtaszczyk and J. B. Avalos,
%Influence of Hydrodynamic Interactions
%on the Kinetics of Colloidal Particles? Adsorption
{\it Phys. Rev. Lett.}, {\bf 80}, 754 (1998).


\bibitem{Tanaka2}
H. Tanaka,
%Roles of hydrodynamic interactions in structure formation of soft matter: protein folding
%as an example
{J. Phys. Cond. Matt.}, {\bf 17}, S2795, (2005).

\bibitem{Baumketner}
A. Baumketner and Y. Hiwatari,
%Influence of hydrodynamic interaction on kinetics and thermodynamics of minimal protein
%models
{\it J. Phys. Soc. Jap.}, {\bf 71}, 3069 (2002).

\bibitem{Vazquez}
M. Carrion-Vazquez, H. B. Li, H. Lu, P. E. Marszalek, A. F. Oberhauser,
and J. M. Fernandez,
% The mechanical stability of ubiquitin is linkage dependent.
{Nat. Struct. Biol.} {\bf 10}, 738, (2003).

\bibitem{Chyan}
C.-L. Chyan, F.-C. Lin, H. Peng, J.-M. Yuan, C.-H. Chang, S.-H. Lin and G. Yang,
% Reversible Mechanical Unfolding of Single Ubiquitin Molecules
{\it Biophys J.}, {\bf 87}, 3995 (2004).


\bibitem{FernandezLi}
J. M. Fernandez and H. Li,
%Force-Clamp Spectroscopy Monitors the Folding Trajectory of a Single Protein,
{\it Science} {\bf 303} 1674 (2004).

\bibitem{Schlierf}
M. Schlierf, H. Li and J. M. Fernandez,
%The unfolding kinetics of ubiquitin captured with
%single-molecule force-clamp techniques
{\it Proc. Natl. Acad. Sci. (USA)} {\bf 101}, 7299, (2004).


\bibitem{Makarov}
D. Makarov,
%Ubiquitin-like protein domains show high resistance to mechanical unfolding
{\it J. Phys. Chem. B.}, {\bf 108}, 745 (2004).

\bibitem{Makarov2}
P-C. Li and D. Makarov,
%Simulation of the mechanical unfolding of ubiquitin
{\it J. Chem. Phys.}, {\bf 121}, 4826 (2004).

\bibitem{cieplakmarszalek}
M. Cieplak and P. E. Marszalek,
%Mechanical unfolding of ubiquitin molecules,
{\it J. Chem. Phys. } {\bf 123} 194903 (2005).


\bibitem{Irback}
A. Irb\"{a}ck, S. Mitternacht, and S. Mohanty,
%Dissecting the mechanical unfolding of ubiquitin
{\it Proc. Natl. Acad. Sci. (USA)} {\bf 102}, 13427 (2005).

\bibitem{cs1}
 P. Szymczak and M. Cieplak,
{\it J. Phys.: Condens. Matter} {\bf 18}, L21, (2006).

\bibitem{West}
 D. K. West, D. J. Brockwell, P. D. Olmsted, S. R. Radford and E. Paci,
{\it Biophys. J.}, {\bf 90}, 287, (2006).


\bibitem{Goabe}
H. Abe and N. Go,
{\it Biopolymers} {\bf 20} 1013 (1981);
%\bibitem{Stakada}
S. Takada,
{\it Proc. Natl. Acad. Sci. (USA)} {\bf 96} 11698 (1999).

\bibitem{ciepho}
M. Cieplak, T. X. Hoang and M. O. Robbins,
%Thermal effects in stretching of Go-like models of titin and
%secondary structures.
{\it Proteins: Struct. Funct. Bio.} {\bf 56} 285 (2004).

\bibitem{sulkowska}
J. I. Su{\l}kowska and M. Cieplak, {\it J. Phys. Cond. Matt.}
(submitted).

\bibitem{Schulten}
H. Lu and K. Schulten,
%Steered molecular dynamics simulation of conformational changes of
%immunoglobulin domain I27 interprete atomic force microscopy observations
{\it Chem. Phys.} {\bf 247}, 141 (1999).

\bibitem{Tsai}
J. Tsai, R. Taylor,  C. Chothia, and M. Gerstein,
%The packing density in proteins: Standard radii and volumes.
{\it J. Mol. Biol.} {\bf 290} 253 (1999).

\bibitem{Pastore}
M. Cieplak, A. Pastore and T. X. Hoang,
%Mechanical properties of the domains of titin in a Go-like model.
{\it J. Chem. Phys.} {\bf 122} 054906 (2004).


\bibitem{veit} T. Veitshans, D. Klimov, and D. Thirumalai,
% Protein folding kinetics:Timescales,
%pathways and energy landscapes
%in terms of sequence-dependent properties.
{\it Folding and Design}, {\bf 2} 1 (1997).

\bibitem{Hoang}
T. X. Hoang and M. Cieplak,
%Sequencing of folding events in Go-like proteins,
{\it J. Chem. Phys.}, {\bf 113}, 8319 (2000).

\bibitem{Lorentz}
H. A. Lorentz,
{\it Lectures on Theoretical Physics}, Macmillan, London, 1927.

\bibitem{Mazo}
R. M. Mazo,
{\it Brownian Motion. Fluctuations, Dynamics, and Applications},
Oxford Science, Oxford, 2002.


\bibitem{Min}
W. Min, G. Luo, B. J. Cherayil, S. C. Kou, and X.S. Xie,
%Observation of a Power-Law Memory Kernel for Fluctuations
%within a Single Protein Molecule
{Phys. Rev. Lett}, {\bf 94}, 198302 (2005).

\bibitem{ermak}
D. L. Ermak and J. A. McCammon,
% Brownian dynamics with hydrodynamic interactions
{\it J. Chem. Phys.}, {\bf 69}, 1352, (1978).

\bibitem{ansell} G. C. Anselle, E. Dickinson, and M. Ludvigsen,
%Brownian Dynamics of Colloidal-aggregate Rotation and
%Dissociation in Shear Flow
{\it J. Chem. Soc., Faraday Trans.} {\bf 81}, 1269 (1985).

\bibitem{Kim}
S. Kim and S. J. Karrila,
{\it Microhydrodynamics: Principles and Selected Applications},
Butterworth-Heinemann, Boston, 1991.


\bibitem{Mazur}
P. Mazur and W. van Saarlos,
% Many-sphere hydrodynamic interactions and mobilities in a suspension.
{\it Physica A}, {\bf 115}, 21 (1982).

\bibitem{Brady}
L. Durlofsky, J. F. Brady, and G. Bossis,
% Dynamic simulation of hydrodynamically
% interacting particles.
{\it J. Fluid Mech.}, {\bf 180} 21 (1987).

\bibitem{Felderhof}
B. U. Felderhof,
% Many-body hydrodynamic interactions in suspensions.
{\it Physica A}, {\bf 151}, 16 (1988).

\bibitem{Ladd}
A. J. C. Ladd,
%Hydrodynamic interactions in a suspension of spherical particles.
{\it  J. Chem. Phys.}, {\bf 88}, 5051, (1988).

\bibitem{Cichocki}
B. Cichocki, B. U. Felderhof, K. Hinsen, E. Wajnryb, and J. Blawzdziewicz,
% Friction and mobility of many spheres in Stokes flow.
{\it J. Chem. Phys.}, {\bf 100}, 3780, (1994).

\bibitem{Brady2}
T. N. Phung, J.F. Brady, and G. Bossis,
% Stokesian Dynamics Simulation of Brownian Suspensions.
{\it J. Fluid Mech.}, {\bf 313}, 181, (1996).

\bibitem{Rotne}
J. Rotne and S. Prager,
%Variational treatment of hydrodynamic interaction on polymers.
{\it J. Chem. Phys.} {\bf 50}, 4831, (1969).

\bibitem{Yamakawa}
H Yamakawa,
%Transport properties of polymer chains in
%dilute solutions. Hydrodynamic interaction.
{\it J. Chem. Phys.} {\bf 53} 436 (1970).

\bibitem{Wajnryb}
E. Wajnryb, P. Szymczak, B. Cichocki,
%Brownian dynamics: divergence of mobility tensor,
Physica A, {\bf 335}, 339, (2004).

\bibitem{Naegele}
G. N\"{a}gele
in {\it Computational Condensed Matter Physics}, ed.: S. Bl\"{u}gel, G. Gompper,
E. Koch, H. M\"{u}ller-Krumbhaar, R. Spatschek, R. G. Winkler, Forschungzentrum
J\"{u}lich, B4:1, 2006.


\bibitem{cs2}
 P. Szymczak and M. Cieplak,
{\it J. Chem. Phys}, {\bf 125}, 164903, (2006).


\bibitem{lemak1} A. Lemak, J. R. Lepock, and J. Z. Y. Chen,
Proteins: Structure, Function and Genetics {\bf 51}, 224 (2003).

\bibitem{lemak2} A. Lemak, J. R. Lepock, J. Z. Y. Chen,
%Unfolding a protein in external fields:
%can we always observe folding immediate states?,
Phys. Rev. E {\bf 67}, 031910 (2003).

\bibitem{Rzehak} R. Rzehak, W. Kromen, T. Kawakatsu and W. Zimmermann,
%Deformation of a Tethered Polymer in Uniform Flow,
Europ. Phys. J. E {\bf 2}, 3 (2000).

\end{thebibliography}
\end{document}